\documentclass[preprint,aps]{revtex4}

\usepackage{lineno}
\usepackage{dcolumn}

\usepackage{bm}
\usepackage{amsmath}
\usepackage{multirow}
\usepackage{graphicx}
\usepackage{float}
\usepackage[colorlinks=true]{hyperref}

\begin{document}

\title{Non-linear quantum effects in electromagnetic radiation \\ of a vortex electron}

\author{D.\,V.~Karlovets}


\affiliation{Tomsk State University, Lenina Ave. 36, 634050 Tomsk, Russia}

\author{A.\,M.~Pupasov-Maksimov}


\affiliation{Universidade Federal de Juiz de Fora, Brasil} 




\date{\today}

\begin{abstract}
There is a controversy of how to interpret interactions of electrons with a large spatial coherence with light and matter.
When such an electron emits a photon, it can do so either as if its charge were confined to a point within a coherence length, 
the region where a square modulus of a wave function $|\psi|^2$ is localized, or as a continuous cloud of space charge spread over it. 
This problem was addressed in a recent study R.~Remez, et al., Phys. Rev. Lett. {\bf 123}, 060401 (2019) where a conclusion was drawn in favor of the first (point) interpretation.
Here we argue that there is an alternative explanation for the measurements reported in that paper, which relies on purely classical arguments and does not allow one to refute 
the second interpretation. We propose an experiment of Smith-Purcell radiation from a non-relativistic vortex electron 
carrying orbital angular momentum, which can unambiguously lead to the opposite conclusion. Beyond the paraxial approximation, the vortex packet has a non-point electric quadrupole moment, 
which grows as the packet spreads and results in a non-linear $L^3$-growth of the radiation intensity with the length $L$ of the grating 
when $L$ is much larger than the packet's Rayleigh length. Such a non-linear effect has never been observed for single electrons and, if detected,
it would be a hallmark of the non-point nature of charge in a wave packet. Thus, two views on $|\psi|^2$ are complementary to each other 
and an electron radiates either as a point charge or as a continuous charge flow depending on the experimental conditions and on its quantum state.
Our conclusions hold for a large class of non-Gaussian packets and emission processes for which the radiation formation length can exceed the Rayleigh length, 
such as Cherenkov radiation, transition radiation, diffraction radiation, and so forth.
\end{abstract}

\maketitle

\section{Introduction}

The particle-wave duality underpinned by de Broglie \cite{dB} lies in the core of quantum mechanics. Modern electron microscopes utilize beams whose transverse coherence length can exceed $1$ mm 
and in a single-particle regime -- for currents lower than 50 nA -- the wave nature of individual electrons is expected to reveal itself in electromagnetic radiation generated during the interaction 
with matter and light. However, it was found in a recent study \cite{PRL19} that optical Smith-Purcell radiation \cite{SP} of electrons with a transverse coherence length $\sigma_{\perp}^{(e)}$ 
larger than 33 $\mu$m occurs as if the charge were confined to a point within this length where a square modulus of a wave function $|\psi|^2$ is localized.
Similar conclusions were also drawn in Ref.\cite{HD} for photoemission in a laser wave, while dependence on the electron packet's size was shown to appear 
when the photons are in the coherent state \cite{Gover} or when the electron's state is different from a simplified plane-wave \cite{Sokolov, Akhiezer, Akhiezer2, Bagrov, Bagrov2, DiPiazza},
especially when an electron Wigner function \cite{Wigner} is not everywhere positive \cite{HD}. The results of Ref.\cite{PRL19} seem to refute a wave-like interpretation of $|\psi|^2$ 
according to which the charge $e$ is spread continuously over the entire coherence length akin to a multi-particle beam. 
On a more fundamental level, the latter interpretation is due to corrections to the classical radiation intensity that arise because of the quantum character of the electron motion 
and are neglected in such quasi-classical approaches as, for instance, an operator method \cite{Baier, BLP}.

Here we show that there is an alternative explanation for the measurements reported in Ref.\cite{PRL19}, which is based on a purely classical concept 
of the so-called \textit{pre-wave zone} \cite{Verz, SPRPW, Mono} and, therefore, it does not allow one to conclude in favor of one of the intepretations. 
We 
demonstrate how to modify the experimental scheme in order to come to the opposite (continuous current density) conclusion without an alternative classical explanation. 
Namely, we propose to use the vortex electrons carying orbital angular momentum (OAM) $\hbar \ell$ \cite{Bliokh} to generate Smith-Purcell radiation. 
Such electrons -- unlike the customary Gaussian beams -- have an intrinsic electric quadrupole moment beyond a paraxial approximation \cite{PRA191, Silenko}, 
which is proportional to the packet's coherence length and the wider the packet is the larger the quadrupole contribution to the radiation. 
Spreading of a non-relativistic vortex packet during its propagation next to a grating can result in a non-linear $L^3$-dependence of the radiation intensity 
on the grating length $L$ due to the quadrupole moment.

The non-linear effects have previously been known only for Smith-Purcell radiation from high-current beams, starting from 1 mA \cite{Andrews1, Andrews2}, or for electrons exposed to a laser field \cite{T}, but never for a single freely propagating electron. Here we predict a non-linear enhancement of the quantum corrections to the classical radiation intensity for a single vortex electron or, more generally, for any non-Gaussian packet with a quadrupole moment, which is also the case for an Airy beam \cite{Airy}, for a Schr\"odinger's cat state, etc. We argue that for the available beams with $\ell \gg 1$ such a non-paraxial quantum effect can be detected and it would be a hallmark of the non-point nature of charge in a wave packet, especially when the recoil is vanishing. Importantly, our conclusions hold for a wide class of emission processes for which the radiation formation length can exceed the packet's Rayleigh length, such as transition radiation, diffraction radiation, emission in a laser pulse, and so on. 
A system of units $\hbar = c = 1$ is used.

\section{Pre-wave zone effects in radiation}

Smith-Purcell radiation as a special case of diffraction radiation \cite{Kaz, SP, Pap, Br, Mono, JETP11} arises as the field of an electron induces a time-varying current density ${\bm j}$ on a grating. 
Quantum-mechanically, the radiation arises due to elastic scattering of a virtual photon by the grating. 
The transverse coherence length of the virtual photon emitted by the electron is
\begin{eqnarray}
\displaystyle \sigma_{\perp}^{(\gamma)} \approx \beta\gamma\lambda \lesssim \lambda\ \text{for}\ \beta \approx 0.4-0.7,
\label{1}
\end{eqnarray}
where $\gamma = \varepsilon/m = 1/\sqrt{1-\beta^2} \gtrsim 1$.
There are at least two reasons why a non-relativistic electron with a large transverse coherence length 
\begin{eqnarray}
& \displaystyle
\sigma_{\perp}^{(e)} \gg \lambda \gtrsim \sigma_{\perp}^{(\gamma)}
\label{2}
\end{eqnarray}
emits Smith-Purcell (diffraction) radiation like a point particle confined inside a region of the width $\sigma_{\perp}^{(e)}$ where $|\psi|^2$ is localized
and not like a cloud of space charge $e$ spread over this region: \textit{(i)} as the radiation is due to scattering of the virtual photons, 
\textit{a radiation formation width} is of the order of $\sigma_{\perp}^{(\gamma)}$, 
not the entire region of $\sigma_{\perp}^{(e)}$, which is profoundly different from radiation by an accelerated electron; 
\textit{(ii)} if a detector is placed at a far distance, $r \gg \sigma_{\perp}^{(e)}$,
a multipole expansion of the radiation intensity holds,
\begin{eqnarray}
& \displaystyle
dW = dW_e + dW_{e\mu} + dW_{eQ} + dW_{\mu} + dW_{Q} + ...,
\label{Wmult}
\end{eqnarray}
even if the packet is wide. 
Here $dW_e$ is due to the electron charge $e$, $dW_{e\mu}$ describes interference of the waves emitted by the charge and by the electron's point magnetic moment\footnote{We neglect the spin. The electron's electric dipole moment is prohibited in the Standard Model \cite{BLP}.} ${\bm \mu}$, $dW_{eQ}$ is due to a \textit{nonpoint} electric quadrupole moment $Q_{ij}$, etc.
In a linear approximation, suitable for currents lower than 1 mA, these multipole moments are coupled to those of the wave packet itself (see the Appendix). 
A key observation here is that all the higher moments \textit{are vanishing} if the packet is Gaussian, at least approximately \cite{PRA191}. That is why, whatever width a packet has it always radiates like a point charge, $dW = dW_e$, within the paraxial approximation. 

\begin{figure}[t]
\centering
\includegraphics[width=.7\linewidth]{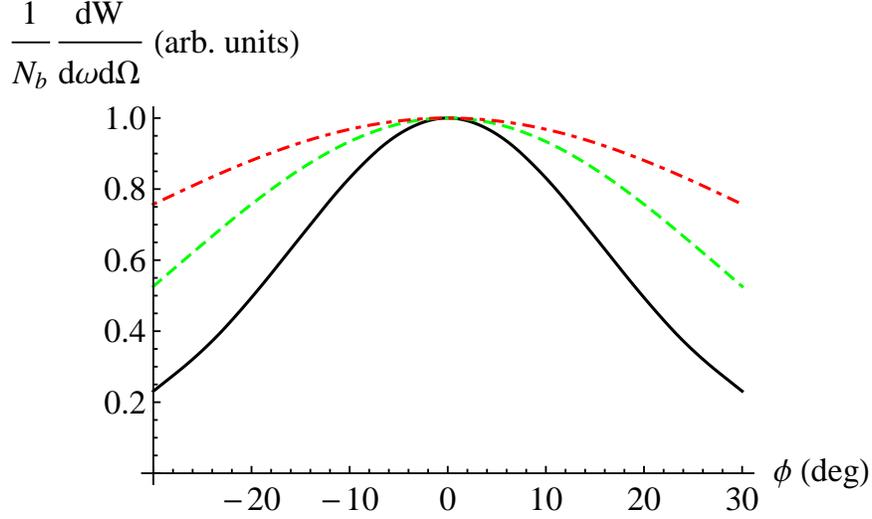}
\caption{Azimuthal distributions of Smith-Purcell radiation for $\lambda = d$, the parameters of Ref.\cite{PRL19} 
and different distances to the detector according to Eq.(\ref{Wb}) and the model \cite{PLA}. 
The green dashed line ($r = 0.5\, r_{\text{p-w}}$) and the red dash-dotted line ($r = 0.3\, r_{\text{p-w}}$) correspond to the pre-wave zone, 
while the black solid line corresponds to the wave zone ($r \gg r_{\text{p-w}}$).}
\label{Fig1}
\end{figure}

Thus, the conclusions of Ref.\cite{PRL19} could have been expected for the chosen experimental conditions but they do not allow one to unambiguously refute 
the continuous current density interpretation because the measurements could support it if the conditions were different. 
Before we formulate them, we demonstrate how the observed in Ref.\cite{PRL19} wide azimuthal distributions can be explained 
by using a classical concept of the pre-wave zone \cite{Verz, Mono, SPRPW}. First, the models of Smith-Purcell radiation from a point charge (see, for instance, \cite{Pap, Br, Mono, JETP11}) 
predict the far-field azimuthal distributions that \textit{are much narrower} than those in Fig.3 of Ref.\cite{PRL19}, see the black solid line in our Fig.\ref{Fig1}. 
This width is a function of the particle energy due to the envelope $dW_e \propto \exp\left\{-\frac{4\pi h}{\beta\gamma\lambda}\,\sqrt{1 + \beta^2\gamma^2\cos^2\Phi\sin^2\Theta}\right\}$ 
where $h$ is an impact-parameter. 
The wide distributions may be a hallmark that the measurements were performed in the pre-wave zone, not in the far field. 

When collecting many photons emitted by many electrons, a transverse region of the grating, 
which participates in the formation of radiation, is of the order of the beam width $\sigma_{\text{b}}^{(e)}$, 
which is much larger than the width of a packet $\sigma_{\perp}^{(e)}$. So, the condition of the wave zone in a plane $\Theta \approx \Phi \approx \pi/2$ (see Fig.2) is \cite{SPRPW}
\begin{eqnarray}
& \displaystyle
r \gg r_{\text{p-w}} = (\sigma_{\text{b}}^{(e)})^2/\lambda.
\label{wz}
\end{eqnarray}
For parameters of Ref.\cite{PRL19}, the pre-wave zone radius $r_{\text{p-w}}$ is found to be
\begin{eqnarray}
& \displaystyle
r_{\text{p-w}} \approx 15\ \text{cm},\,\, \sigma_{\text{b}}^{(e)} = 300\, \mu\text{m},\cr
& \displaystyle
r_{\text{p-w}} \approx 6.7\ \text{m},\,\, \sigma_{\text{b}}^{(e)} = 2\, \text{mm}.
\label{rpwz}
\end{eqnarray}
Thus, the measurements of Ref.\cite{PRL19} are likely to have taken place in the pre-wave zone where the azimuthal distributions must be very broad \cite{SPRPW}.

To take this effect into account, one needs to average the one-particle intensity, $dW^{\text{class}}({\bf r}_T)$, not with $|\psi|^2$ as in Eq.(4) of Ref.\cite{PRL19}
but with a beam transverse distribution function $\rho_b({\bf r}_T)$,
\begin{eqnarray}
\displaystyle \frac{dW}{d\omega d\Omega} = \int d^2{\bf r}_T\, \rho_b ({\bf r}_T)\,\frac{dW^{\text{class}}({\bf r}_T)}{d\omega d\Omega}.
\label{Wb}
\end{eqnarray}
The function $\rho_b$ can be Gaussian, $\rho_b \propto N_b \exp\{-{\bf r}_T^2/2(\sigma_b^{(e)})^2\} $, normalized to a number $N_b$ of electrons in the beam. 
Importantly, both Eq.(\ref{Wb}) and Eq.(4) of Ref.\cite{PRL19} indirectly imply that the detector can be placed in the pre-wave zone because the far-field intensity 
does not depend on the transverse shift ${\bf r}_T$ at all. Indeed, this shift is a phase rotation, $\psi({\bf p}) \to \psi({\bf p})\, e^{-i{\bf p}\cdot{\bf r}_T}$, 
and the intensity \cite{BLP} $dW^{\text{far-field}}/d\omega d\Omega = -\frac{e^2 \omega^2}{(2\pi)^2}\,
{j_{fi}}_{\mu}(k) \left({j_{fi}}^{\mu}(k)\right)^*,\, j^{\mu}_{fi}(k) = \int d^4x\, \bar{\psi}_f(x)\gamma^{\mu}\psi_i(x)\,e^{ikx}$, stays invariant under it.
Unlike Eq.(\ref{Wb}), the wave zone formula deals with the momentum representation, which is quite natural -- see the Appendix.

To calculate $dW^{\text{class}}({\bf r}_T)$ at an arbitrary distance $r$ we use the model of Ref.\cite{PLA}, although the azimuthal distributions 
are largely model-independent. As can be seen in Fig.\ref{Fig1}, the green and red lines fit the data in Fig.3 of Ref.\cite{PRL19} much better than the far-field line does,
which represents an alternative classical explanation of the unusually wide distributions reported in Ref.\cite{PRL19}.

\section{Smith-Purcell radiation from a vortex electron}

Now we are going to propose an experiment in which an electron radiates in the far field as if it had its charge spread over the entire region of $\sigma_{\perp}^{(e)}$, 
while the contribution, which depends on the coherence length, is \textit{non-linearly enhanced}. For non-Gaussian packets there appear additional terms in Eq.(\ref{Wmult}) 
because the far-field intensity $dW$ 
is generally sensitive to the size of the electron packet and to its shape defined by the phase $\varphi({\bf p})$ of the wave function (see the Appendix). The vortex electrons with OAM $\ell$ \cite{Bliokh}, 
the Airy beams \cite{Airy}, as well as superpositions of states can serve as such non-Gaussian packets and they also have an electric quadrupole moment, which 
-- unlike the magnetic moment -- has a finite radius defined by the packet's coherence length. Importantly, the quadrupole contribution comes about only \textit{beyond the paraxial approximation} \cite{PRA18}, 
which implies that the packet is narrow (unlike that of Ref.\cite{PRL19}) and the OAM is large $\ell \gg 1$.

\begin{figure}[t]
 \centering
 \includegraphics[width=.7\linewidth]{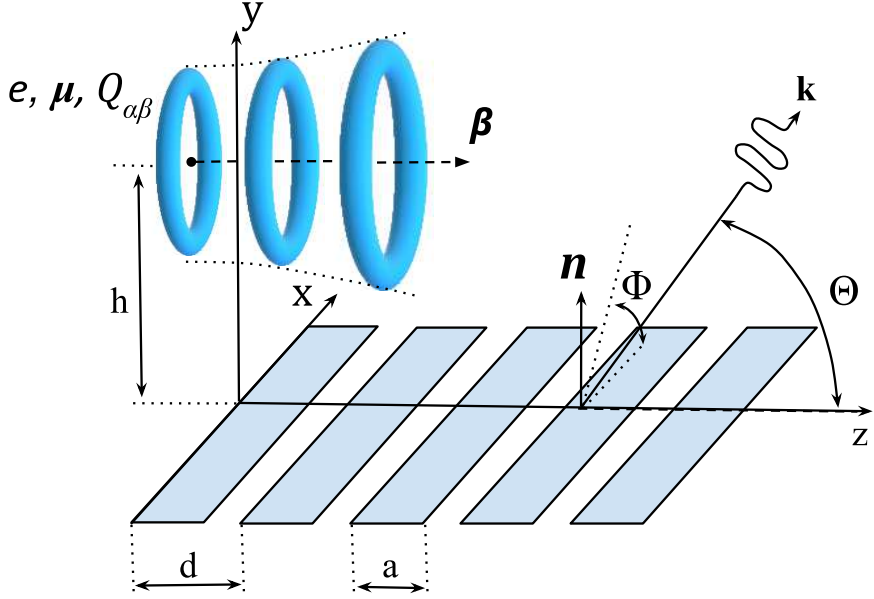}
 \caption{Smith-Purcell radiation of a vortex electron packet possessing a point charge $e$, a point magnetic moment ${\bm \mu}$, and a non-point electric quadrupole moment $Q_{\alpha\beta}(t)$,
which grows as the packet spreads. The radiation wavelength is $\lambda = d\, (\beta^{-1} - \cos\Theta)/n,\, n=1,2,3,...$}
 \label{Fig2}
\end{figure}

Consider Smith-Purcell radiation generated by a non-relativistic vortex electron, see Fig.2. Its magnetic moment and the electric quadrupole moment are \cite{PRA191, PRA192, Silenko}
\begin{eqnarray}
\displaystyle {\bm \mu} = {\hat{\bm z}}\,\frac{\ell}{2m},\quad Q_{ij}(t) = (\sigma_{\perp}^{(e)}(t))^2\, \text{diag}\{1/2,1/2,-1\},
\label{Moments}
\end{eqnarray}
where $\sigma_{\perp}^{(e)}(t) = \sigma_{\perp}^{(e)}\,\sqrt{1 + t^2/t_d^2}$, $t_d = (t_c/|\ell|)\, (\sigma_{\perp}^{(e)}/\lambda_c)^2$ is a spreading time, and $\lambda_c = 1/m \approx 3.9 \times 10^{-11}\, \text{cm}$ is the Compton wavelength. Neglecting both the recoil and the quadratic corrections $dW_{\mu}$, $dW_{Q}$, etc., we have the following radiation intensity:
\begin{eqnarray}
\displaystyle
dW = dW_e + dW_{e\mu} + dW_{eQ},
\label{SerShort}
\end{eqnarray}
where neither $dW_e$ nor $dW_{e\mu}$ depends on the packet's width, but $dW_{eQ}$ does. We calculate these terms according to the model \cite{PLA} in which the surface current is induced 
by the given fields ${\bf E}_e, {\bf E}_{\mu}, {\bf E}_Q$ of the first three moments of the vortex electron derived in Ref.\cite{PRA192}. 
The leading (classical) term $dW_{e}$ is defined by Eqs.(57),(58) in Ref.\cite{JETP11}, while the magnetic moment contribution, $dW_{e\mu}/dW_{e} \sim \ell\, \cos\Phi\, \lambda_c/\lambda$, 
can reach $10^{-4}$ for $\ell \sim 10^3$ and $\lambda \sim 1\, \mu$m but it vanishes at $\Phi = \pi/2$ due to the symmetry considerations \cite{PRL13}.

Importantly, both the corrections to $dW_e$ in Eq.(\ref{SerShort}) have a quantum origin. While $dW_{e\mu}$ is $\ell$ times larger than the recoil (i.e., $\ell\, \omega/\varepsilon \gg \omega/\varepsilon$, see Ref.\cite{PRL13}), the term $dW_{eQ}$ is due to quantum character of the trajectory \cite{Bagrov, Bagrov2, Sokolov, BLP}, which is also supposed to be larger than the recoil. Such ``geometric'' corrections can be noticeable for the emisson of a coherent superposition of packets with a non-everywhere positive Wigner function \cite{HD, DiPiazza}. However, as we show hereafter, they can also be non-linearly enhanced due to the spreading, while the recoil stays vanishing, $\omega \ll \varepsilon$, which can take place even for a single-electron state with an everywhere positive Wigner function.

When $\sigma_{\perp}^{(e)} \ll \lambda$, the quadrupole contribution consists of two parts, $dW_{eQ} = dW_{eQ_1}(N) + dW_{eQ_2}(N^3)$. The former represents a standard non-paraxial correction \cite{PRA18, PRA192},
\begin{eqnarray}
& \displaystyle dW_{eQ_1}/dW_{e} \sim \ell^2 \frac{\lambda_c^2}{(\sigma_{\perp}^{(e)})^2},
\label{nonpar1}
\end{eqnarray}
while the latter part is due to the spreading (the term $t^2/t_d^2 \equiv \langle z\rangle^2/z_R^2$) and it can be neglected for relativistic electrons \cite{Bagrov, BLP, Sokolov} or when the radiation formation length is smaller than the Rayleigh length $z_R = \beta t_d = \beta\,(\lambda_c/|\ell|)\,(\sigma_{\perp}^{(e)})^2/\lambda_c^2$. For non-relativistic energies, however, the Rayleigh length does not exceed a few cm for relevant parameters and the spreading can noticeably modify the radiation if the length $L$ of the grating of $N$ strips is large: $L = N d \gg z_R$. In this case, the quadrupole contribution integrated over frequencies for the first diffraction order $n=1$ is found as
\begin{eqnarray}
& \displaystyle \frac{dW_{eQ_2}}{d\Omega} \approx N^2\, \ell^2 \frac{\lambda_c^2}{(\sigma_{\perp}^{(e)})^2}\, \frac{2\pi^2}{3\beta^4\gamma^4}\,\frac{d^2}{\lambda^2(\Theta)}\,\, \frac{dW_{e}}{d\Omega},\cr
& \displaystyle dW_{eQ_2}/dW_{e} \sim N^2\, \ell^2 \frac{\lambda_c^2}{(\sigma_{\perp}^{(e)})^2}.
\label{nonpar2}
\end{eqnarray}
where $\lambda(\Theta) = d\, (\beta^{-1} - \cos\Theta)$. For a long grating, $N \gg 1$, this ratio can be only moderately attenuated, $dW_{eQ_2}/dW_{e} \lesssim 1$, while both the ordinary non-paraxial contribution and the recoil can still be small, $dW_{eQ_1}/dW_{e} \ll 1, \omega/\varepsilon \ll 1$. 

Most importantly, while the classical intensity $dW_e$ linearly grows with the number of strips $N$, the non-paraxial contribution $dW_{eQ_2}$ grows \textit{non-linearly}\footnote{It comes about due to the integration of the spreading term along the grating, $\int dz\, z^2/z_R^2 \propto (N d)^3/z_R^2$.}, as $N^3$. 
This remarkable feature is a direct consequence of the delocalized nature of charge in the twisted packet 
and it puts an upper limit on the grating length $L_{\text{max}} = N_{\text{max}} d$ for which the radiation losses stay small compared to the particle's energy.
This limit can be derived by demanding that both the recoil and the quadratic corrections can be neglected, $\omega/\varepsilon \sim \lambda_c/\lambda \ll dW_{eQ_2}/dW_{e} \ll 1$, which yields
\begin{eqnarray}
& \displaystyle \sqrt{\frac{\lambda_c}{\lambda}}\,\frac{\sigma_{\perp}^{(e)}}{\lambda_c |\ell|}  \ll N \ll \frac{\sigma_{\perp}^{(e)}}{\lambda_c |\ell|}.
\label{Ineq}
\end{eqnarray}
For the moderately large OAM, $|\ell| \sim 10-100$, and $\sigma_{\perp}^{(e)} \sim 1\,\text{nm} - 1\,\mu\text{m}$, we have $\sigma_{\perp}^{(e)}/\lambda_c |\ell| \sim 10-10^5$,
so the number $N_{\text{max}}$ can be taken as $0.1-0.2$ of this value. Note that in contrast to the magnetic moment effects \cite{PRL13}, 
the observation of this non-linear enhancement does not necessarily require as large an OAM as possible.

\begin{figure}[t]
 \centering
 \includegraphics[width=.7\linewidth]{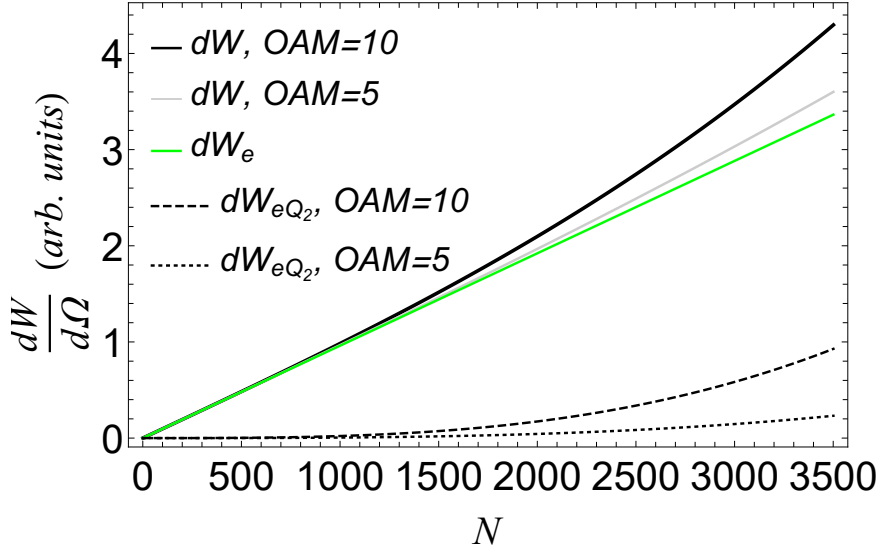}
 \caption{Dependence of Smith-Purcell radiation on the number of strips for $d = 10\,\mu$m, $\sigma_{\perp}^{(e)} = 100$ nm, $h = 2.7\,\mu$m, $z_R \approx 1.3$ mm, $N_{\text{max}} \approx 3500$, $L_{\text{max}} \approx 3.5$ cm, $\Theta = \Phi = \pi/2$. While for a point charge this dependence is linear (the green line), a non-point vortex packet with a quadrupole moment reveals an $N^3$ dependence for $Nd \gg z_R$.}
 \label{Fig3}
\end{figure}

\begin{figure}[t]
 \centering
 \includegraphics[width=.7\linewidth]{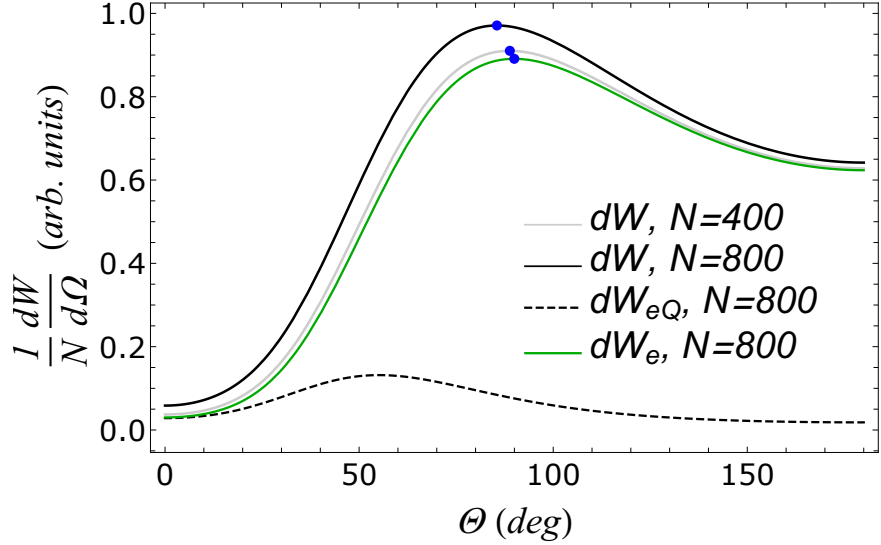}
 \caption{Polar dependence of Smith-Purcell radiation for $d = 100\,\mu$m, $\sigma_{\perp}^{(e)} = 20$ nm, $\ell = 10$, $h = 33\,\mu$m, $z_R \approx 70\,\mu$m, $N_{\text{max}} \approx 800$, $L_{\text{max}} \approx 8$ cm, $\Phi = \pi/2$. The maximum (the blue dot) is shifted due to the quadrupole contribution.}
 \label{Fig4}
\end{figure}

The easiest way to detect this non-linear effect is to perform measurements in the perpendicular plane, at $\Theta=\Phi=\pi/2$, and to compare the radiation 
from at least three gratings of different length. In this geometry, the magnetic moment term vanishes, $dW_{e\mu} = 0$, and $dW_{eQ_2}$ can reach some $10-20$\% of the leading term $dW_e$. 
The effect can more easily be detected in IR and THz ranges, for which the grating period should be larger than $10\, \mu$m. 
In Fig.\ref{Fig3} we present the non-linear growth of the intensity with the number of strips, which can be seen with a naked eye, while in Fig.\ref{Fig4} the enhancement for the small polar angles, 
$dW_{eQ_2}(\Theta=0)/dW_{eQ_2}(\Theta=\pi/2) \approx 4$, is shown accompanied with a several-degree shift of the maximum.
If detected, this shift could also serve as an evidence of the quadrupole contribution. Note that for very wide packets, $\sigma_{\perp}^{(e)} \gg \lambda$, the quadratic corrections $dW_{\mu}, dW_Q$ 
and higher-order terms can become important, which is why we do not consider the case $\sigma_{\perp}^{(e)} \gtrsim 33\, \mu$m of Ref.\cite{PRL19}.

As the electron coherence length in a vicinity of a cathode does not exceed a few nm \cite{Cho, Eh} and for vortex packets it scales as $\sigma_{\perp}^{(e)} \propto \sqrt{|\ell|}$,
the grating must be placed not too far from the vortex electron source or, alternatively, the focusing can be applied. 
When detecting many photons from electrons of a beam, it is important to have the beam angular divergence as small as possible, 
otherwise many electrons could hit the grating well before they reach the part where the quadrupole contribution becomes noticeable.
For the optical range and the grating period $d=416$ nm, the maximal grating length $L_{\text{max}} \sim 10\, \mu$m
matches the effective interaction length of the beam used in Ref.\cite{PRL19} (the distance before an electron hits the grating) 
for $\sigma_{\perp}^{(e)} \sim 10$ nm and $|\ell| \sim 200$, which seems feasible, although the beam focusing could be needed.
Instead of minimizing the beam divergence, one could also rotate the grating so that to minimize the electron losses, although at the expense of statistics.

The above non-linear enhancement can also reveal itself in other processes with the non-relativistic non-Gaussian packets for which the radiation formation length 
can be much larger than the Rayleigh length, such as Cherenkov radiation and diffraction radiation in a cylindrical channel of a finite length, 
transition radiation in a slab, Compton emission in a laser pulse, and so on.





\section{Conclusion}



Concluding, we have argued that the classical pre-wave zone effect could have been the reason for the wide azimuthal distributions of Smith-Purcell radiation reported in Ref.\cite{PRL19}.
The continuous current density interpretation of the wave function can still be used when the radiation intensity depends on the electron coherence length,
which is generally the case \cite{BLP, Sokolov, Bagrov, Bagrov2}. We have predicted a non-linear enhancement of the quantum non-paraxial corrections
to the classical radiation intensity due to the non-local nature of charge in a spreading packet of the vortex electron. 
Moreover, any non-Gaussian packet with an electric quadrupole moment can emit radiation in the far-field as if its charge were spread 
over the entire coherence length. This non-point contribution can reveal itself in a non-linear growth of the intensity for a family of emission processes 
when the radiation formation length exceeds the Rayleigh length. Our findings support Bohr's complementarity principle and demonstrate 
that a choice between the two seemingly contradictory interpretations of a square modulus of the wave function depends on the experimental conditions 
-- in particular, on a distance to the detector -- and on a quantum state and energy of the projectile.



We are grateful to A.\,Aryshev, V.\,G.~Bagrov, A.\,P.~Potylitsyn, and, especially, to P.\,O.~Kazinski and A.\,A.~Tishchenko for fruitful discussions and criticism. 
This work is supported by the Russian Science Foundation (Project No.\,17-72-20013).

\section{Appendix: radiation from an electron wave packet}

\subsection{Generalities} 

Consider radiation of a charged wave packet either in external electromagnetic field or when interacting with a medium in the lowest order of the perturbation theory in quantum electrodynamics (QED).
The formula (3) of the paper is based on a multipole expansion of the transition current density ${\bf j}_{fi}$, in which two quantum effects are present: \textit{(i)} the recoil and \textit{(ii)} the effects of the electron wave-packet's size and shape. The possibility of such a multipole expansion not only in classical electrodynamics but also in QED 
follows from linearity of the latter on the tree-level. Indeed, the radiation intensity of the classical current $j_{\mu}(x)$ \textit{in the far-field} 
is given by Eq.(14.70) in Ref.\cite{J}, which can be written as follows:
\begin{eqnarray}
& \displaystyle \frac{dW}{d\omega d\Omega} = -\frac{\omega^2}{(2\pi)^2}\, j_{\mu}(k) \left(j^{\mu}(k)\right)^*,\quad j_{\mu}(k) = \int d^4x\, j_{\mu}(x)\, e^{it\omega - i{\bf k}{\bf x}},
\label{EqJ}
\end{eqnarray}
when integrating over all space and time. A probability to emit a photon by an electron in the lowest order of QED is
\begin{eqnarray}
& \displaystyle d\nu = |S_{fi}|^2\, \frac{d^3 k}{(2\pi)^3},\, S_{fi} = -ie \int d^4 x\, j^{\mu}_{fi}(x)\, A_{\mu}^*(x),\cr
& \displaystyle j^{\mu}_{fi}(x) = \bar{\psi}_f(x)\gamma^{\mu}\psi_i(x).
\label{S2}
\end{eqnarray}
When the photon is detected in the wave zone as a plane wave with $A_{\mu}(x) = \frac{\sqrt{4\pi}}{\sqrt{2\omega}}\, e_{\mu}(k)\, \exp\{-it\omega + i {\bf k}{\bf x}\}$,
the radiated energy summed over the photon polarizations by $e_{\mu}e_{\nu}^*\to - g_{\mu\nu}$ is found as
\begin{eqnarray}
& \displaystyle \frac{dW}{d\omega d\Omega} = \omega\,\frac{d\nu}{d\omega d\Omega} = -\frac{\omega^2}{(2\pi)^2}\, e^2{j_{fi}}_{\mu}(k) \left({j_{fi}}^{\mu}(k)\right)^*.
\label{sWq}
\end{eqnarray}
The only difference from Eq.(\ref{EqJ}) is that the electron final state does not coincide with its initial state, while both these states are arbitrary and are not necessarily plane waves.
This correspondence is a manifestation of the Bohr's complementarity principle and it is because of this that the general quantum formulas for radiation intensity look similar to those of the classical electrodynamics, see § 45 in \cite{BLP}. This is in particular the case for such a wide class of processes as polarization radiation beyond the dipole approximation, including diffraction and Smith-Purcell radiation, in which the quantrum recoil is vanishing but the multipole structure of the current is retained, which means that the ``geometric'' corrections due to the size and shape of the packet are taken into account (see below).

The contributions of higher multipole moments are described in classical electrodynamics by keeping higher-order terms in expansion of the Green's function into series. Analogously, the mutipole expansion of the radiation intensity in QED can be obtained by expanding the plane-wave component $\exp\{-i{\bf k}{\bf x}\}$ of the final photon into series over the spherical waves -- see § 46, 47 in \cite{BLP}. Such a multipole expansion holds irrespective of the specific emission process, also when the recoil is vanishing, which is implied in Eq.(3) of the main text. As a result, the matrix element $S_{fi}$ in (\ref{S2}) will represent a series over the multipole contributions and the intensity will look like Eq.(3). However, as we show hereafter, for this expansion to have a sense it is important that the current $j^{\mu}_{fi}(x)$ be spatially localized, which means that both the initial and final electrons are described as wave packets rather than plane waves.


\subsection{Role of the size and shape of the electron packet} 


Now we are going to demonstrate how to take into account the size and shape of an electron wave packet in radiation or the shape of a beam in \textit{incoherent} radiation of $N_b$ electrons, 
which is typical in an electron microscope. Let the initial electron be described as an arbitrary packet with a wave function being a superposition of plane waves,
\begin{eqnarray}
& \displaystyle \psi_i(x) = \int\frac{d^3p}{(2\pi)^3}\, \psi ({\bf p})\, \frac{u_i(p)}{\sqrt{2\varepsilon}}\, e^{-it\varepsilon + i{\bf p}{\bf x}},\ \bar{u}_i(p)u_i(p) = 2m,\ \varepsilon = \sqrt{{\bf p}^2 + m^2},\cr
& \displaystyle \int d^3 x |\psi_i(x)|^2 = \int \frac{d^3p}{(2\pi)^3} |\psi({\bf p})|^2 = 1.
\label{psiin}
\end{eqnarray}
The matrix element and the probability to emit a plane-wave photon become
\begin{eqnarray}
& \displaystyle S_{fi} = \int\frac{d^3p}{(2\pi)^3}\, \psi ({\bf p})\, S_{fi}^{\text{(pw)}}({\bf p}),\cr 
& \displaystyle d\nu = \int \frac{d^3p}{(2\pi)^3}\frac{d^3p^{\prime}}{(2\pi)^3}\,\psi({\bf p}) \psi^*({\bf p}^{\prime})\,S_{fi}^{\text{(pw)}}({\bf p})\,(S_{fi}^{\text{(pw)}})^*({\bf p}^{\prime})\,\frac{d^3k}{(2\pi)^3} = \cr 
& \displaystyle = \int \frac{d^3p}{(2\pi)^3}\frac{d^3q^{\prime}}{(2\pi)^3}\,\psi({\bf p} + {\bf q}/2) \psi^*({\bf p} - {\bf q}/2)\,S_{fi}^{\text{(pw)}}({\bf p} + {\bf q}/2)\,(S_{fi}^{\text{(pw)}})^*({\bf p} - {\bf q}/2)\,\frac{d^3k}{(2\pi)^3},
\label{Snpw}
\end{eqnarray}
where we use the new variables 
\begin{eqnarray}
& \displaystyle
({\bf p},{\bf p}^{\prime})\to ({\bf p} + {\bf q}/2,{\bf p} - {\bf q}/2).
\label{WigVar}
\end{eqnarray} 
If we deal with a single electron and not with a multi-particle beam, one can completely neglect the dependence of $S_{fi}^{\text{(pw)}}$ on ${\bf q}$, which is called \textit{the paraxial approximation}. 
The corrections due to small ${\bf q}$ arise beyond the paraxial regime because of the \textit{auto-correlation} of the scattering amplitude or due to its phase $\zeta_{fi}$:
$$
S_{fi}^{\text{(pw)}}({\bf p}) = |S_{fi}^{\text{(pw)}}({\bf p})|\, e^{i\zeta_{fi}({\bf p})}.
$$
If the phase is constant, which also depends on the final electron state, the corrections vanish exactly.
For a beam, the leading term with $|S_{fi}^{\text{(pw)}}({\bf p})|^2$ describes the incoherent emission of uncorrelated particles, 
while the first correction due to non-vanishing ${\bf q}$ takes inter-particle correlations (coherence effects) into account.
The small-${\bf q}$ expansion of $S_{fi}^{\text{(pw)}}$ is justified because the electron wave packet is normalized and, therefore, the function $\psi({\bf p})$
can behave at large ${\bf p} \to \infty$, for instance, as $\psi({\bf p}) \propto \exp\left\{-({\bf p} - \langle {\bf p}\rangle)^2/2(\delta p)^2\right\}$. Then 
\begin{eqnarray}
& \displaystyle
\psi^*({\bf p} - {\bf q}/2) \psi({\bf p} + {\bf q}/2) \propto  \exp\left\{-({\bf p} - \langle {\bf p}\rangle)^2/(\delta p)^2 - {\bf q}^2/(2\delta p)^2\right\}
\label{qs}
\end{eqnarray} 
at large ${\bf p}$.

The leading term in the paraxial approximation is thus
\begin{eqnarray}
\displaystyle d\nu^{(\text{incoh})} = \int \frac{d^3p}{(2\pi)^3}\frac{d^3q^{\prime}}{(2\pi)^3}\,\psi({\bf p} + {\bf q}/2) \psi^*({\bf p} - {\bf q}/2)\cr
\displaystyle \times |S_{fi}^{\text{(pw)}}({\bf p})|^2\frac{d^3k}{(2\pi)^3} = \int \frac{d^3p}{(2\pi)^3}\, n ({\bf 0}, {\bf p}, 0)\,d\nu^{\text{(pw)}}({\bf p}),
\label{Snpw0}
\end{eqnarray} 
or for the radiation intensity in the wave zone (cf. Eq.(3), (4) in Ref.\cite{HD})
\begin{eqnarray}
\displaystyle \frac{dW^{(\text{incoh})}}{d\omega d\Omega} = \int \frac{d^3p}{(2\pi)^3}\, n ({\bf 0}, {\bf p}, 0)\,\frac{dW^{\text{(pw)}}({\bf p})}{d\omega d\Omega},
\label{Snpw01}
\end{eqnarray}
where we have used the definition of a Wigner function \cite{Wigner},
\begin{eqnarray}
& \displaystyle n({\bf x}, {\bf p}, t) = \int\ \frac{d^3q}{(2\pi)^3}\, \psi^*({\bf p} - {\bf q}/2, t) \psi({\bf p} + {\bf q}/2, t)\, e^{i{\bf q}{{\bf x}}},\cr 
& \displaystyle \psi({\bf p}, t) = \psi({\bf p})\,e^{-it\varepsilon({\bf p})}.
\label{Wn}
\end{eqnarray}
The formula (\ref{Snpw01}) allows one to exactly take into account the spatial shape and width of the radiating packet because the momentum uncertainty $\delta p$ is connected with the former 
as $\sigma_{\perp}^{(e)} = 1/\delta p$. Importantly, it is only for a Gaussian packet that the Wigner function $n ({\bf 0}, {\bf p}, 0)$ coincides with $|\psi({\bf p})|^2$ (cf. Eq.(3) in Ref.\cite{HD}),
while for a vortex electron, for instance, it does not -- cf. Eq.(68) in Ref.\cite{PRD}. Thus, for non-Gaussian electron packets equations (\ref{Snpw0}), (\ref{Snpw01}) also depend on a phase $\varphi({\bf p})$ of the electron wave function 
\begin{eqnarray}
\displaystyle
\psi({\bf p}) = |\psi({\bf p})|\, e^{i\varphi({\bf p})}
\label{WVP}
\end{eqnarray} 
and they are applicable for packets with the not-everywhere positive Wigner functions -- say, Schrödinger's cat states, coherent superpositions of vortex states, etc. 

The main difference of Eq.(\ref{Snpw01}) from Eq.(4) in Ref.\cite{PRL19} is that the former uses the momentum representation, while the latter -- the coordinate one.
The use of the momentum representation is natural and even \textit{unavoidable} for the wave zone because the radiation source is completely delocalized,
which is why one has to deal with momenta, not coordinates. As clearly seen from Eq.(\ref{S2}), the far-field radiation probability does not depend
on the transverse shift ${\bf r}_T$ of the radiating electron because such a shift changes only the phase of both the initial and final electrons as 
$$
\psi_{i,f}({\bf p}) \to \psi_{i,f}({\bf p})\, e^{-i{\bf p}\cdot{\bf r}_T},
$$
to which the intensity is not sensitive. 
The intensity is sensitive, however, to a phase rotation of the initial electron alone, $\psi({\bf p}) \to \psi({\bf p})\, e^{i\varphi(\bf{p})}$,
which is why the higher multipole moments can make a non-vanishing contribution to the far-field. We would like to emphasize that the quantum state of the final electron is \textit{not specified} here and the final photon is described as a delocalized plane wave, which means that the photon is detected in the wave zone. If the final electron were also described as a plane wave, which means that it is not detected, 
the radiation intensity would \textit{not} depend on the phase $\varphi({\bf p})$ of the initial electron when integrating over all space and time from $-\infty$ to $+\infty$.
Such a phase dependence takes place only if the final electron is also described as a spatialy localized wave packet, which means that it is detected at a certain distance (not too far) from the radiation region.
It is this case, which is the most natural for comparison with the classical theory because the transition current $j^{\mu}_{fi}(x)$ is spatially localized, while for the plane-wave final electron 
it is not so and, therefore, the wave zone \textit{cannot} be defined\footnote{The lack of the dependence on the phase $\varphi({\bf p})$ when both the final particles are the plane waves looks very natural: this phase defines the shape of the incoming packet. But if the current $j^{\mu}_{fi}(x)$ is not localized such a shape does not have a sense.}.



For emission of many photons by a beam of electrons, the Wigner function is normalized to a number $N_b$ of particles in the beam,
$$
\int \frac{d^3p}{(2\pi)^3}\, n ({\bf 0}, {\bf p}, 0) = N_b.
$$
In this case, Eq.(\ref{Snpw01}) describes incoherent radiation, which is a good approximation for small radiation wavelengths $\lambda \ll \sigma_{b}^{(e)}$ 
and the low-current (single-electron) regime, typical for electron microscopes. The opposite case of $\lambda \gtrsim \sigma_{b}^{(e)}$ and the bunched electrons 
can be realized in a particle accelerator, for which the leading term (\ref{Snpw01}) is no longer sufficient, see Ref.\cite{KS}. 

On the contrary, to describe the radiation in the pre-wave zone it is natural to use the coordinate representation. 
The corresponding classical formula is given by Eq.(6) in the main text of the paper,
\begin{eqnarray}
\displaystyle \frac{dW}{d\omega d\Omega} = \int d^2{\bf r}_T\, \rho_b ({\bf r}_T)\,\frac{dW^{\text{class}}({\bf r}_T)}{d\omega d\Omega}.
\label{WbApp}
\end{eqnarray}
When the detector is in the far field, the dependence of $dW^{\text{class}}$ on ${\bf r}_T$ vanishes and we are left with
\begin{eqnarray}
\displaystyle \frac{dW^{\text{far-field}}}{d\omega d\Omega} = N_b\,\frac{dW^{\text{class}}}{d\omega d\Omega},
\label{WbApp2}
\end{eqnarray}
which reflects the well-known fact that an incoherent form-factor for a beam equals unity \cite{KS, TS}. As has been recently shown in Ref.\cite{TS}, 
the incoherent form-factor \textit{can differ from unity} when the grating in Smith-Purcell radiation or a target in transition and diffraction radiation is spatially limited 
-- say, when the grating has a width smaller than the transverse coherence length of the virtual photon $\beta\gamma\lambda$, so the radiation formation width is defined
by the geometrical sizes of the target. 

Analogously, the pre-wave effect also comes about due to the finite radiation formation width but because the detector is moved closer to the target. 
Eq.(\ref{WbApp}) explicitly demonstrates, therefore, that the incoherent form-factor \textit{also differs from unity} for the radiation in the pre-wave zone.
In this sense, the wide azimuthal distributions measured in Ref.\cite{PRL19} can be treated as an evidence of such a form-factor. 
This conclusion holds not only for Smith-Purcell radiation, but also for a much wider class of emission processes, including diffraction radiation, Cherenkov radiation, transition radiation, 
Compton and Thomson scattering in laser fields, and so forth.

\subsection{The quasi-classical regime of emission by an electron packet} 

Now we are going to demonstrate how to study emission when the quantum effects are small and treated as corrections to the classical formula. Along with the recoil, these corrections depend on the shape and size of the electron wave packet via the multipole expansion. We emphasize that both these effects are inherently quantum, 
so the separation of them in radiation intensity is a rather delicate task even for the Gaussian packets -- see, for instance, \cite{Bagrov, Bagrov2, DiPiazza}.  However when the quantum numbers defining the shape of a non-Gaussian packet are large (say, the orbital angular momentum for a vortex electron $\ell \gg 1$), the emission is always \textit{quasi-classical} \cite{BLP} and one can neglect the spin contribution ($\mathcal O(\omega/\varepsilon)$) compared to the contributions originating from the non-Gaussianity of the packet (say, for the vortex packet it is of the order of $\ell\, \omega/\varepsilon \gg \omega/\varepsilon$ \cite{PRL13}).

We start again with the general matrix element 
\begin{eqnarray}
& \displaystyle S_{fi} = -ie \int d^4 x\, j^{\mu}_{fi}(x)\, A_{\mu}^*(x),\quad j^{\mu}_{fi}(x) = \bar{\psi}_f(x)\gamma^{\mu}\psi_i(x),
\label{Sgen}
\end{eqnarray}
and take both the incoming electron and the final electron as some wave packets
,
\begin{eqnarray}
& \displaystyle \psi_i(x) = \int\frac{d^3p}{(2\pi)^3}\, \psi_i ({\bf p})\, \frac{u_i(p)}{\sqrt{2\varepsilon_i}}\, e^{-it\varepsilon + i{\bf p}{\bf x}},\ \bar{u}_i(p)u_i(p) = 2m,\cr
& \displaystyle \psi_f(x) = \int\frac{d^3p}{(2\pi)^3}\, \psi_f ({\bf p})\, \frac{u_f(p)}{\sqrt{2\varepsilon_f}}\, e^{-it\varepsilon_f + i{\bf p}{\bf x}},\ \bar{u}_f(p)u_f(p) = 2m.
\label{inout}
\end{eqnarray}
Depending on the external field, these packets can be coherent superpositions of the Volkov states in a plane wave, of the Landau states in magnetic field, etc. 
The transition current looks as follows:
\begin{eqnarray}
& \displaystyle j^{\mu}_{fi}(x) = \int \frac{d^3 p}{(2\pi)^3}\frac{d^3 p_f}{(2\pi)^3}\, \psi^*_f({\bf p}_f) \psi_i({\bf p})\, \frac{\bar{u}_f({\bf p}_f)}{\sqrt{2\varepsilon_f}}\,\gamma^{\mu} \frac{u_i({\bf p})}{\sqrt{2\varepsilon_i}}\, e^{-ix (p-p_f)} = \cr
& \displaystyle = \int \frac{d^3 p}{(2\pi)^3}\frac{d^3 q}{(2\pi)^3}\, \frac{\psi^*_f({\bf p} - {\bf q}/2)}{\sqrt{2\varepsilon({\bf p} - {\bf q}/2)}} \frac{\psi_i({\bf p} + {\bf q}/2)}{\sqrt{2\varepsilon({\bf p} + {\bf q}/2)}}\, \bar{u}_f({\bf p} - {\bf q}/2)\,\gamma^{\mu} u_i({\bf p} + {\bf q}/2)\cr
& \displaystyle \times \exp\left\{-it (\varepsilon ({\bf p} + {\bf q}/2) - \varepsilon ({\bf p} - {\bf q}/2)) + i{\bf x} {\bf q}\right\},
\label{jgen}
\end{eqnarray}
where we again use the variables (\ref{WigVar}) and no approximations are used at this stage. The indices $i$ and $f$ denote all the rest quantum numbers the packets can possess (spin, orbital angular momentum, etc.)

Now we notice that the variable ${\bf q} = {\bf p} - {\bf p}_f$ is a momentum transfer for each plane-wave component composing the wave packets.
The large values $|{\bf q}| \gg \delta p$ are suppressed in the quasi-classical case $f\to i$ analogously to Eq.(\ref{qs}). However, even in the general quantum regime
the large momentum transfers are attenuated by the rapidly oscillating exponent $\exp\{i{\bf x} {\bf q}\}$. So the effective values of ${\bf q}$ are
\begin{eqnarray}
& \displaystyle |{\bf q}| \lesssim 1/|{\bf x}| \sim 1/\sigma_{\perp} = \delta p,
\label{qeff}
\end{eqnarray}
whatever shape the packets have. 

An expansion of the bispinors into series over ${\bf q}$ yields ($i,f$ are just spin indices here)
\begin{eqnarray}
&& \displaystyle \bar{u}_f({\bf p} - {\bf q}/2)\,\gamma^{\mu} u_i({\bf p} + {\bf q}/2) = \bar{u}_f({\bf p})\gamma^{\mu} u_i({\bf p}) + \cr
&& \displaystyle + \frac{{\bf q}}{2}\,\left(\bar{u}_f({\bf p})\gamma^{\mu}\frac{\partial u_i({\bf p})}{\partial {\bf p}} - \frac{\partial \bar{u}_f({\bf p})}{\partial {\bf p}}\gamma^{\mu} u_i({\bf p})\right ) + \mathcal O({\bf q}^2).
\label{bisps}
\end{eqnarray}
The first correction \textit{due to recoil} here depends on the electron spin\footnote{It is clear from the Gordon identity, $\bar{u}({\bf p} - {\bf q}/2)\,\gamma^{\mu} u({\bf p} + {\bf q}/2) = \frac{1}{2m}\,\bar{u}({\bf p} - {\bf q}/2)\,(2p^{\mu} - i\sigma^{\mu\nu}q_{\nu}) u({\bf p} + {\bf q}/2)$.} ${\bm \zeta}$ and in the quasi-classical regime with $f\to i$ (no spin-flip) it looks as follows (we omit the index $i$) \cite{PRA18}:
\begin{eqnarray}
&& \displaystyle \bar{u}({\bf p})\gamma^{\mu}\frac{\partial u({\bf p})}{\partial p_j} - \frac{\partial \bar{u}({\bf p})}{\partial p_j}\gamma^{\mu} u({\bf p}) = 2i\left\{\frac{1}{\varepsilon + m}\, [{\bm \zeta}\times {\bf p}]_j, \frac{p_j}{\varepsilon}\,\frac{{\bm \zeta}\times {\bf p}}{\varepsilon + m} + {\bf e}_j \times {\bm \zeta}, \right\},
\label{bispsqc}
\end{eqnarray}
where ${\bf e}_j$ is a unit vector along the $\text{j}^{\text{th}}$ axis. So, this correction is generally attenuated as $\omega/\varepsilon \ll 1$,
it coincides with the corresponding term in Eq.(2.4) of Ref.\cite{Bagrov}, and vanishes for an unpolarized electron. Therefore, for an unpolarized electron we have simply
\begin{eqnarray}
&& \displaystyle \bar{u}_f({\bf p} - {\bf q}/2)\,\gamma^{\mu} u_i({\bf p} + {\bf q}/2) \to 2 p^{\mu} = 2m u^{\mu},
\label{bispsunp}
\end{eqnarray}
even if the recoil is not vanishing. In this case, the integral over ${\bf q}$ in (\ref{jgen}) yields a following function:
\begin{eqnarray}
& \displaystyle \tilde{n}({\bf x}, {\bf p}, t) = \int\ \frac{d^3q}{(2\pi)^3}\, \frac{\psi^*({\bf p} - {\bf q}/2, t)}{\sqrt{2\varepsilon({\bf p} - {\bf q}/2)}} \frac{\psi({\bf p} + {\bf q}/2, t)}{\sqrt{2\varepsilon({\bf p} + {\bf q}/2)}}\, e^{i{\bf q}{{\bf x}}},
\label{Wnmod}
\end{eqnarray}
which is very similar to the electron Wigner function Eq.(\ref{Wn}), but transforms differently under the Lorentz boosts. So the current for an polarized electron looks like
\begin{eqnarray}
& \displaystyle j^{\mu}_{f\to i}(x) = \int \frac{d^3 p}{(2\pi)^3}\,2 p^{\mu}\, \tilde{n}({\bf x}, {\bf p}, t)
\label{jftoi}
\end{eqnarray}
and depends on the electron phase $\varphi$. We stress that this current is \textit{not fully classical} because the quantum recoil and the packet's phase are taken into account, 
but the rest quantum numbers (say, orbital angular momentum) do not change during the radiation.

Let us now analyse effects of the packet's shape and size for an unpolarized electron. The former are defined by the phase $\varphi$, while the latter arise due to the finite momentum width $\delta p = 1/\sigma_{\perp}^{(e)} \equiv 1/\sigma_{\perp}$. Let us first denote 
\begin{eqnarray}
& \displaystyle \Psi({\bf p}) = \frac{\psi({\bf p})}{\sqrt{2\varepsilon({\bf p})}}.
\label{Psi}
\end{eqnarray}
Then we represent the new wave functions according to Eq.(\ref{WVP}) and find
\begin{eqnarray}
& \displaystyle \Psi^*({\bf p} - {\bf q}/2) \Psi({\bf p} + {\bf q}/2) = \left(|\Psi|^2 + \frac{1}{4}\, q_i q_j \left(|\Psi|\frac{\partial^2|\Psi|}{\partial p_i \partial p_j} - \left(\frac{\partial|\Psi|}{\partial p_i}\right)\left(\frac{\partial|\Psi|}{\partial p_j}\right)\right) + \mathcal O(q^4)\right)\cr
& \displaystyle \times \exp\left\{i{\bf q} \frac{\partial \varphi}{\partial {\bf p}} + \mathcal O(q^3)\right\}.
\label{psiexp1}
\end{eqnarray}
where $\Psi \equiv \Psi({\bf p}), \varphi \equiv \varphi({\bf p})$.
The exponent here is due to electric and magnetic \textit{dipole moments} of the packet. The mean value of the former is \cite{PRA191}
\begin{eqnarray}
& \displaystyle
{\bf d} = -\left\langle\frac{\partial \varphi({\bf p})}{\partial {\bf p}}\right\rangle.
\label{d}
\end{eqnarray}
However, the true \textit{intrinsic} electric dipole moment of an electron packet is vanishing as it is prohibited by the CPT theorem of the Standard Model.
The mean value of the electric moment (but not of the magnetic one) can be killed by shifting the origin of coordinates or by the choice of initial conditions ${\bf x}_0$ \cite{PRA191}, which implies the following phase rotation:
\begin{eqnarray}
& \displaystyle
\Psi \to \Psi\,\exp\left\{-i{\bf x}_0 {\bf p}\right\},\ {\bf x}_0 = - {\bf d} = \left\langle\frac{\partial \varphi}{\partial {\bf p}}\right\rangle.
\label{din}
\end{eqnarray}
As a result, we have instead
\begin{eqnarray}
& \displaystyle \Psi^*({\bf p} - {\bf q}/2) \Psi({\bf p} + {\bf q}/2) = \left(|\Psi|^2 + \frac{1}{4}\, q_i q_j \left(|\Psi|\frac{\partial^2|\Psi|}{\partial p_i \partial p_j} - \left(\frac{\partial|\Psi|}{\partial p_i}\right)\left(\frac{\partial|\Psi|}{\partial p_j}\right)\right) + \mathcal O(q^4) 
\right)\cr
& \displaystyle \times \exp\left\{i{\bf q} \left(\frac{\partial \varphi}{\partial {\bf p}} - \left\langle\frac{\partial \varphi}{\partial {\bf p}}\right\rangle\right) + \mathcal O(q^3)\right\}.
\label{psiexp2}
\end{eqnarray}
This ambiguity -- that is, dependence of the matrix element on the initial conditions -- is well-known (see, for instance, \cite{Akhiezer, Akhiezer2}) and for a non-Gaussian packet such a choice of the origin guarantees that we work with intrinsic values of the multipole moments. 

One can also expand the energies in the exponent as follows:
\begin{eqnarray}
&& \displaystyle \varepsilon ({\bf p} + {\bf q}/2) - \varepsilon ({\bf p} - {\bf q}/2) = {\bf q} {\bf u} + \mathcal O({\bf q}^3),\ {\bf u} \equiv {\bf u}({\bf p}) = \frac{{\bf p}}{\varepsilon({\bf p})}.
\label{exps}
\end{eqnarray}
After this, the integral over ${\bf q}$ yields a delta-function and the current looks as follows (note that we use both $\psi$ and $\Psi = \psi/\sqrt{2\varepsilon}$ here):
\begin{eqnarray}
& \displaystyle j^{\mu}_{f\to i}(x) = \int \frac{d^3 p}{(2\pi\hbar)^3}\,\frac{p^{\mu}}{\varepsilon}\, \Big (|\psi|^2 + \frac{1}{4} D_{ij}\, \hat{p}_i \hat{p}_j + \mathcal O(\hbar^4)\Big)\,
\delta \left({\bf x} - {\bf u}t + \hbar\left(\frac{\partial \varphi}{\partial {\bf p}} - \left\langle\frac{\partial \varphi}{\partial {\bf p}}\right\rangle\right)\right) \cr
& \displaystyle \equiv \int \frac{d^3 p}{(2\pi\hbar)^3}\, \Big (|\psi|^2 + \frac{1}{4} D_{ij}\, \hat{p}_i \hat{p}_j + \mathcal O(\hbar^4)\Big) j^{\mu}_{\text{quasi-cl.}}({\bf x}, {\bf p}, t; \hbar),\cr
& \displaystyle j^{\mu}_{\text{quasi-cl.}}({\bf x}, {\bf p}, t; \hbar) = \frac{p^{\mu}}{\varepsilon}\,\delta \left({\bf x} - {\bf u}t + \hbar\left(\frac{\partial \varphi}{\partial {\bf p}} - \left\langle\frac{\partial \varphi}{\partial {\bf p}}\right\rangle\right)\right),\cr
& \displaystyle D_{ij} = 2\varepsilon \left(|\Psi|\frac{\partial^2|\Psi|}{\partial p_i \partial p_j} - \left(\frac{\partial|\Psi|}{\partial p_i}\right)\left(\frac{\partial|\Psi|}{\partial p_j}\right)\right),
\label{jftoi2}
\end{eqnarray}
where $\hat{\bf p} = -i\hbar {\bm \nabla}$ and we have restored the Planck's constant $\hbar$. Comparing this with Eq.(\ref{jftoi}), we see that the unpolarized electron's Wigner function is everywhere positive now, eventhough the packet is not Gaussian (cf. \cite{HD}). Treating the term $\mathcal O(\hbar)$ as a perturbation, one can also write this via the fully classical current as follows:
\begin{eqnarray}
& \displaystyle j^{\mu}_{f\to i}(x) = \int \frac{d^3 p}{(2\pi\hbar)^3}\, \Bigg (|\psi|^2 + i |\psi|^2 \left(\frac{\partial \varphi}{\partial {\bf p}} - \left\langle\frac{\partial \varphi}{\partial {\bf p}}\right\rangle\right)\,\hat{\bf p}  + \cr
& \displaystyle + \frac{1}{4}\left(D_{ij} - 2 |\psi|^2 \left(\frac{\partial \varphi}{\partial p_i} - \left\langle\frac{\partial \varphi}{\partial p_i}\right\rangle\right)\left(\frac{\partial \varphi}{\partial p_j} - \left\langle\frac{\partial \varphi}{\partial p_j}\right\rangle\right)\right)\, \hat{p}_i \hat{p}_j + \mathcal O(\hbar^3)\Bigg) j^{\mu}_{\text{cl.}}({\bf x}, {\bf p}, t),\cr
& \displaystyle j^{\mu}_{\text{cl.}}({\bf x}, {\bf p}, t) = \frac{p^{\mu}}{\varepsilon}\,\delta \left({\bf x} - {\bf u}t\right).
\label{jclass}
\end{eqnarray}
Depending on the boundary conditions, the rectilinear motion here corresponds either to Cherenkov radiation or to transition radiation, etc.
A generalization of this for arbitrary classical motion in a given field is obvious, 
$$
{\bf u}t \to {\bf r}(t).
$$

Thus the transition current represents a functional of the classical current and of the classical trajectory \cite{Akhiezer, Akhiezer2, Bagrov} and its quantum corrections due to the recoil are proportional to $\hbar$ and depend on the derivatives of the packet's phase. Remarkably, even when the recoil is vanishing ($\hbar\omega/\varepsilon \to 0$) the current still represents a superposition of trajectories with the different momenta \cite{Akhiezer, Akhiezer2} defined by the wave function $|\Psi|^2$, about which we have not made any assumptions. If this function is, for instance, of a Gaussian form,
\begin{eqnarray}
& \displaystyle
|\Psi|^2 \propto \exp\left\{-\frac{({\bf p} - \langle{\bf p}\rangle)^2}{(\delta p)^2}\right\},
\label{gau}
\end{eqnarray}
the current is equal to
\begin{eqnarray}
& \displaystyle
j^{\mu}_{f\to i}(x) = j^{\mu}_{\text{cl.}}({\bf x}, \langle{\bf p}\rangle, t) + \mathcal O\left(\frac{(\delta p)^2}{m^2}\right),
\label{jclass1}
\end{eqnarray}
and it acquires an inherently quantum \textit{non-paraxial correction} \cite{PRA18} 
\begin{eqnarray}
& \displaystyle
\frac{(\delta p)^2}{m^2} = \frac{\lambda_c^2}{\sigma_{\perp}^2} \ll 1
\label{nonpar}
\end{eqnarray}
due to the packet's finite size $\sigma_{\perp} = \hbar/\delta p$. 
Thus this size can influence the radiation, although only when the packet is very narrow, $\sigma_{\perp} \gtrsim \lambda_c$,
so the ratio $\lambda_c^2/\sigma_{\perp}^2$ does not exceed $10^{-6}$ for relevant parameters. It is these corrections that are neglected in such quasi-classical methods 
as, for instance, the operator method \cite{Baier, BLP} or the eikonal approximation \cite{Akhiezer2}.

The packet's shape, defined by the phase $\varphi$, influences the first quantum correction to the current but not the leading term.
An important exception here, however, is the vortex electrons because for them \cite{PRA18, PRA191}
$$
|\psi|^2 \propto p_{\perp}^{2|\ell|},
$$
and the transition current depends on \textit{the absolute value} of the electron OAM $|\ell|$ already in the leading order, 
which results in an enhancement of the non-paraxial correction (\ref{nonpar}), $\lambda_c^2/\sigma_{\perp}^2 \to |\ell|\,\lambda_c^2/\sigma_{\perp}^2$ \cite{PRA18}.

Let us now derive a more general expression for the current, which allows one to study two opposite limiting cases: \textit{(i)} a delocalized plane-wave electron and \textit{(ii)} a point-like classical one.
We suppose that the electron wave function has a Gaussian envelope, analogously to Eq.(\ref{qs}),
\begin{eqnarray}
& \displaystyle
\psi({\bf p}) = \tilde{\psi}({\bf p})\,\exp\left\{-\frac{({\bf p} - \langle{\bf p}\rangle)^2}{2(\delta p)^2}\right\},\ \tilde{\psi}({\bf p}) = |\tilde{\psi}|\,e^{i\varphi}.
\label{psipre}
\end{eqnarray}
Then instead of the delta-function the integral over ${\bf q}$ yields the following result:
\begin{eqnarray}
& \displaystyle j^{\mu}_{f\to i}(x) = \frac{(\delta p)^3}{\pi^{3/2}\hbar^3}\,\int \frac{d^3 p}{(2\pi\hbar)^3}\, \frac{p^{\mu}}{\varepsilon}\, \Bigg (|\tilde{\psi}|^2 + \frac{(\delta p)^2}{2} \Big(\text{Tr}\,D_{ij} - \cr & \displaystyle - 2 \sigma_{\perp}^{-2} \left({\bf x} - {\bf u}t + \hbar\left(\frac{\partial \varphi}{\partial {\bf p}} - \left\langle\frac{\partial \varphi}{\partial {\bf p}}\right\rangle\right)\right)_i\left({\bf x} - {\bf u}t + \hbar\left(\frac{\partial \varphi}{\partial {\bf p}} - \left\langle\frac{\partial \varphi}{\partial {\bf p}}\right\rangle\right)\right)_j D_{ij}\Big) + \cr 
& \displaystyle + \mathcal O((\delta p)^4) \Bigg)\, \exp\left\{-\frac{({\bf p} - \langle{\bf p}\rangle)^2}{(\delta p)^2} - \sigma_{\perp}^{-2}\left({\bf x} - {\bf u}t + \hbar\left(\frac{\partial \varphi}{\partial {\bf p}} - \left\langle\frac{\partial \varphi}{\partial {\bf p}}\right\rangle\right)\right)^2\right\},
\label{jgen2}
\end{eqnarray}
where $\sigma_{\perp} = \hbar/\delta p = \mathcal O(\hbar)$. When $\sigma_{\perp} \to 0$ (or $\hbar \to 0$) for arbitrary $\delta p$, we return to Eq.(\ref{jclass}). 
However when $\delta p \to 0$ for non-vanishing $\sigma_{\perp}$, we have 
\begin{eqnarray}
& \displaystyle j^{\mu}_{f\to i}(x) = \text{const}\, |\tilde{\psi}(\langle{\bf p}\rangle)|^2\, \frac{\langle p\rangle^{\mu}}{\varepsilon} \equiv \frac{1}{V}\, \frac{\langle p\rangle^{\mu}}{\langle\varepsilon\rangle}\, ,
\label{jpw}
\end{eqnarray}
which is a delocalized current of the plane wave state $\psi(x) = (2\langle\varepsilon\rangle V)^{-1/2}\, u_i(\langle p\rangle) \exp\{-i\langle p\rangle x\}$.

We now return to Eq.(\ref{jclass}) and notice that the first correction due to recoil in the matrix element $S_{fi} =-ie \int d^4 j_{fi}^{\mu}(x) A^*_{\mu}(x)$ looks as follows after the integration by parts (we again omit $\hbar$):
\begin{eqnarray}
& \displaystyle  S_{fi} \propto ie\int d^4x \frac{d^3 p}{(2\pi)^3}\,|\psi|^2\, \left(\frac{\partial \varphi}{\partial {\bf p}} - \left\langle\frac{\partial \varphi}{\partial {\bf p}}\right\rangle\right)\, j^{\mu}_{\text{cl.}}({\bf x}, {\bf p}, t)\, {\bm \nabla} A^*_{\mu}(x)|_{{\bf x} = {\bf u}t}.
\label{1stq}
\end{eqnarray}
Analagously, the second quantum correction contains ${\nabla}_i{\nabla}_j A^*_{\mu}(x)|_{{\bf x} = {\bf u}t}$. Thus we see that the series in powers of the recoil -- that is, $\omega/\varepsilon$ -- automatically generates a multipole series in the matrix element. In particular, the second correction in Eq.(\ref{jclass}) depends on $(\partial \varphi/\partial p_i)(\partial \varphi/\partial p_j)$ and, therefore, on the packet's \textit{electric quadrupole momentum} \cite{PRA191, Silenko}. For a plane-wave photon, we have ${\bm \nabla} A^*_{\mu}(x)|_{{\bf x} = {\bf u}t} = -i{\bf k}A^*_{\mu}(t,{\bf u}t)$. The radiation intensity $dW$ defined by $|S_{fi}|^2$ thus includes the leading classical term and two sorts of quantum corrections: 
\begin{itemize}
\item
Those due to the recoil, which also depend on the packet's multipole moments due to the phase $\varphi$, 
\item
The non-paraxial corrections due to the packet's finite size. 
\end{itemize}
The intensity also depends on the interference between the multipole contributions -- see Eq.(3) in the main text of the paper.
Importantly, these conclusions hold both \textit{(i)} for radiation of an accelerated electron in an external field and \textit{(ii)} when the particle interacts with a medium and no acceleration is required (Cherenkov radiation, transition radiation, Smith-Purcell radiation, etc.).

Finally, note that when the packet's quantum numbers, which define its shape, are large (for instance, $\ell \gg 1$ for a vortex electron), we can neglect the terms of the order of $\omega/\varepsilon$ compared to the contribution from the phase $\varphi$. Say, for a vortex electron we have 
$$
\varphi = \ell \phi
$$
and the first quantum correction to the current and to the matrix element \textit{depends on the sign} of the OAM $\ell$ via $\partial \varphi/\partial {\bf p}$. 
One can retrieve this magnetic dipole contribution by the following asymmetry:
\begin{eqnarray}
& \displaystyle \frac{dW(\ell) - dW(-\ell)}{dW(\ell) + dW(-\ell)} = \frac{dW_{e\mu}}{dW_e} = \mathcal O\left(\ell \frac{\omega}{\varepsilon}\right),
\label{Assym}
\end{eqnarray}
analogously to Ref.\cite{PRL13}, which is $\ell$ times larger than the corresponding spin asymmetry for a Gaussian packet. Likewise, the quadrupole contribution without the spreading is attenuated as $\ell^2\,\lambda_c^2/\sigma_{\perp}^2$ (see Eq.(9) in the main text), which can also be much larger than $\omega/\varepsilon$ for $|\ell| \gg 1$. 
Compared to the quasi-classical regime of emission by relativistic particles \cite{Baier, Akhiezer2}, in which the electron motion is classical but the recoil is kept, here we have the opposite situation:
the recoil is vanishing but the effects due to non-Gaussianity of the electron packet are enhanced. We will present other details elsewhere.

We would like to emphasize that while the above first-order QED approach is applicable not only to processes in the given external fields but also to Cherenkov radiation and transition radiation in a given medium, the Smith-Purcell radiation from a conducting grating cannot, strictly speaking, be described in a model-independent way within the first order of the perturbation theory in QED. This is because we either have to take the incident field of the moving electron as given and to consider radiation of the induced surface current (as in \cite{Kaz, PLA, Mono, Br, JETP11, MI, Sh}) or to take a given surface wave and to consider radiation of the electron in its field (as in \cite{Gover, PRL19}). Clearly, the predictions of both these phenomenological approaches can be different, exactly as they are so in the classical framework. In this paper, we rely on the former (surface current) approach, whose validity was experimentally verified for Smith-Purcell radiation (see, for instance, \cite{Black})
and for other emission processes.

\end{document}